\documentclass[onecolumn]{article}
\usepackage{graphics}
\usepackage[varg]{txfonts} % Times fonts
\usepackage[latin1]{inputenc}
\usepackage{natbib}
\title{Radioactive and kinematic tracers of feedback from massive stars}
\author{Rasmus Voss\thanks{rvoss$@$science.ru.nl}\\
Department of Astrophysics/IMAPP,\\ Radboud University Nijmegen,\\ PO Box 9010, NL-6500 GL Nijmegen, the Netherlands
}
\date{}
\begin{document}
\maketitle
\begin{center}
{\large {\bf Abstract}}\\
The mixing of ejecta from young stars into the interstellar medium is 
an important process in the interplay between star formation and galaxy 
evolution. A unique window into these processes is provided by the 
radioactive isotopes $^{26}$Al, traced by its $\gamma$-ray 
decay lines at 1.8 MeV. With a mean lifetime of $\sim$1 Myr it is a 
long-term tracer 
of nucleosynthesis for massive stars.  Our population synthesis code 
models the ejection of $^{26}$Al, together with the $^{60}$Fe, the kinetic 
energy and UV radiation for a population of massive stars. We have applied 
the code to study the nearby Orion region and the more massive Carina 
region and found good agreement with observational constraints. 
\end{center}

\section{Introduction}
\label{sec:intro}
Feedback from massive stars plays a crucial role in the formation
of stars, as it shapes the ISM and its subsequent star formation activity. 
The main feedback originates from the ejection of matter from massive stars 
through their winds and supernova explosions, and from their intense 
emission at short wavelengths into the UV. This UV radiation creates 
large photoionized regions around the stars, and the kinetic energy 
associated with ejection of stellar matter pushes at the ISM, 
together creating large shells and cavities

We developed a new population synthesis tool to analyze groups of massive stars,
where we model the emission of different forms of energy and matter from the
stars of the association. In particular, the ejection of the two radioactive 
isotopes $^{26}$Al and $^{60}$Fe is followed,
as well as the emission of hydrogen ionizing photons, and the kinetic energy
of the stellar winds and supernova explosions. We investigate various 
alternative astrophysical inputs and the resulting output sensitivities, 
especially effects due to the inclusion of rotation in stellar models.
As the aim of the code is the application
to relatively small populations of massive stars, special care is taken to
address their statistical properties. Our code incorporates both analytical 
statistical methods applicable to small populations, as well as extensive
Monte Carlo simulations. A thorough description of the methods and the
comparisons between different input models can be found in \cite{Voss-popsyn}.

\begin{figure}
\resizebox{0.35\columnwidth}{!}{%
\includegraphics{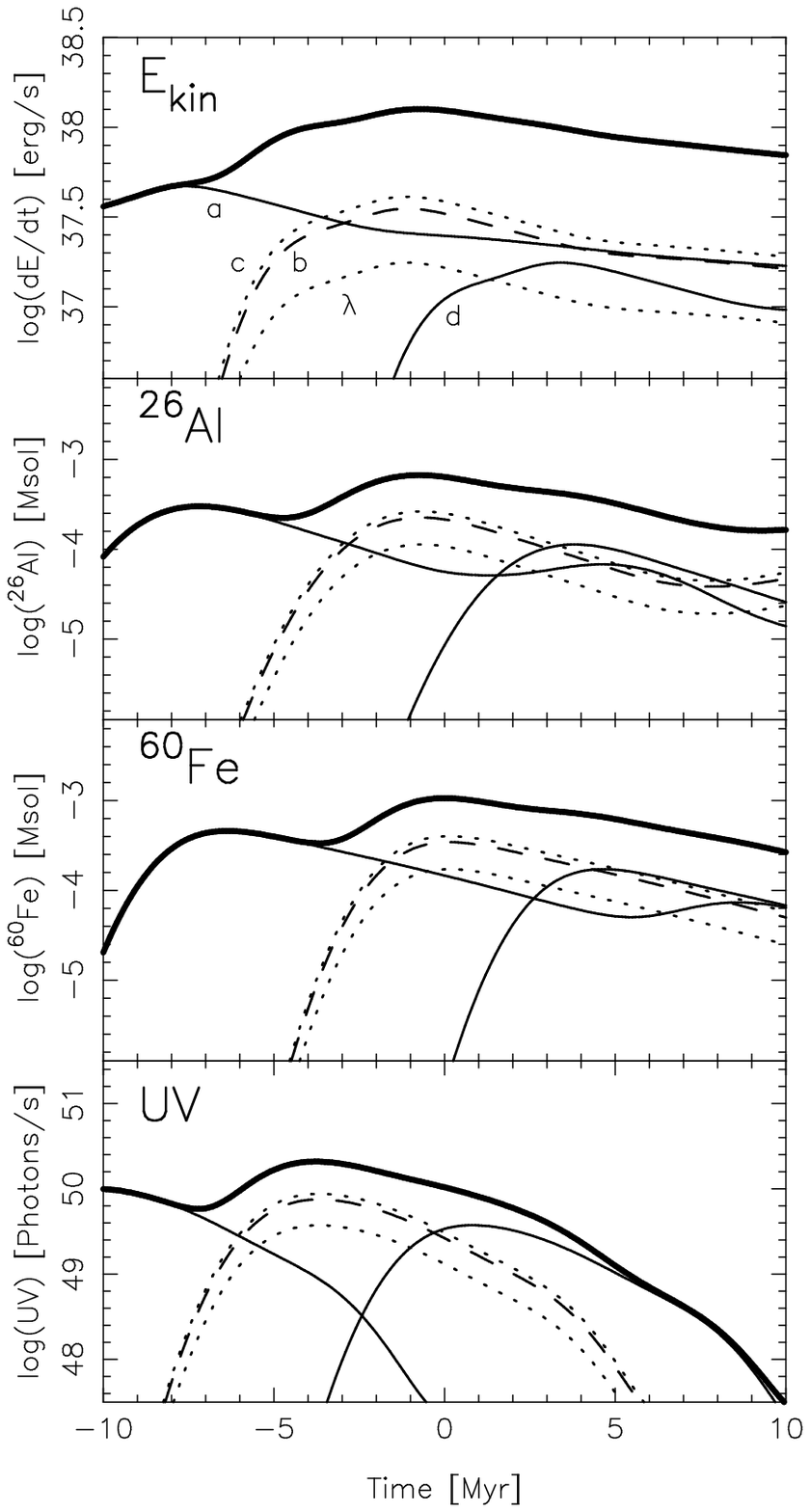} }
\resizebox{0.65\columnwidth}{!}{%
\includegraphics{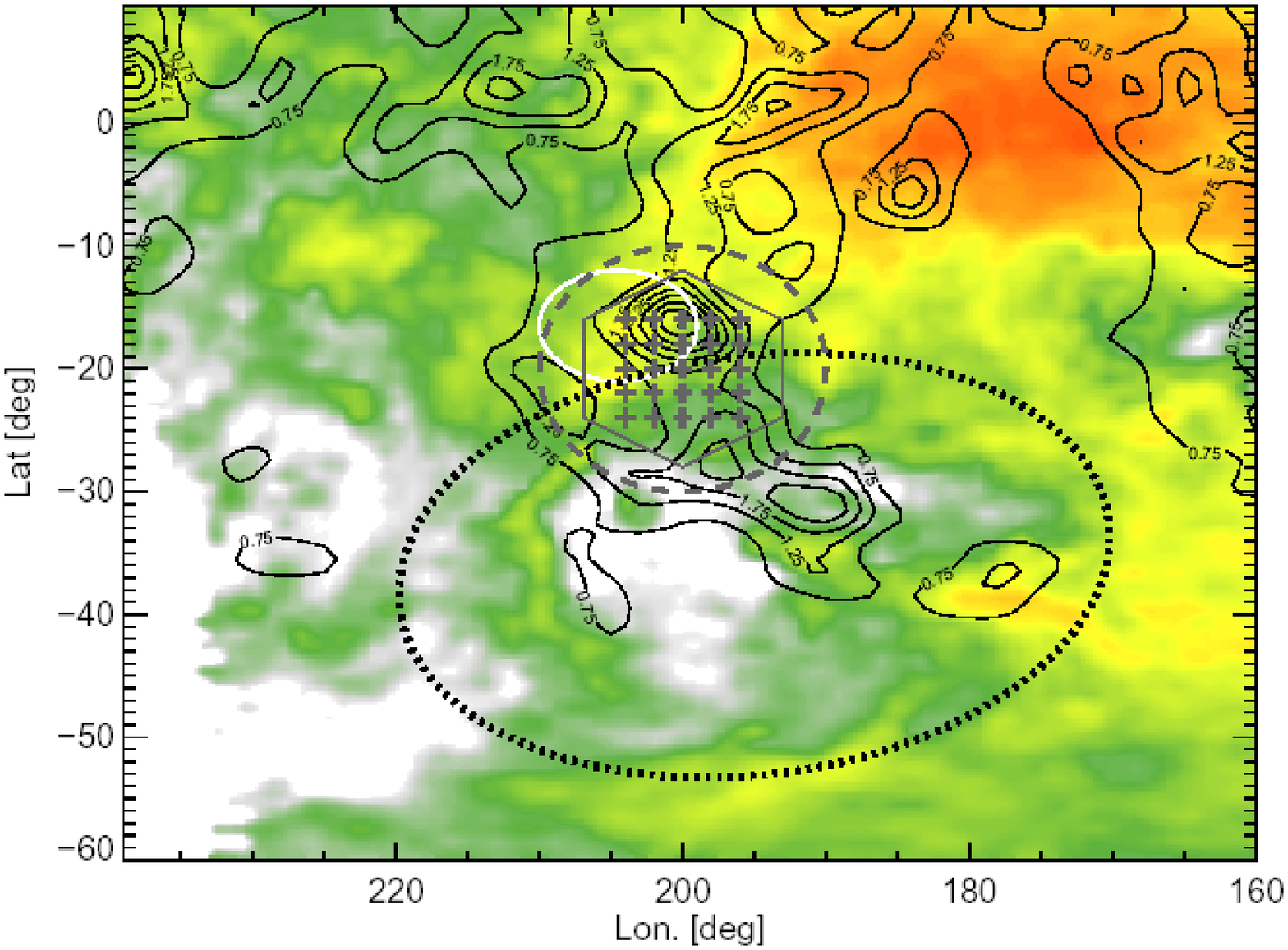} }
\caption{{\bf Left}: Our simulated output of kinetic energy, $^{26}$Al, $^{60}$Fe 
and ionizing radiation from the massive stars in the Orion region. 
The thin lines indicates the outputs from the individual regions of 
the Orion OB1 association. It can be seen that the radioactive elements 
are good tracers of the kinetic energy output, making them useful for 
the study of feedback from massive stars. The results were shown to be 
consistent with observed properties of the region: ISM excavation by the 
kinetic energy, free-free emission from the ionized ISM and gamma-rays 
from the radioactive decay of $^{26}$Al. {\bf Right} The COMPTEL-observed 
1.8 MeV emission (contours), superimposed on a velocity integrated HI map 
of Orion-Eridanus bubble (color image). Orion OB1 is indicated 
(white ellipse), as well as the outer shell of the HI superbubble 
(large, black-dotted ellipse).}
\label{fig:1}       % Give a unique label
\end{figure}

\section{Models of the Orion and Carina regions}
\label{sec:orion}
We have used our population synthesis tool to analyze two regions
with young massive stars, from which the $^{26}$Al signal has been
measured. The Orion OB1 association is located at a distance of
$\sim350-400$ pc and hosts several groups of young stars with ages
0-12 Myr. We have analyzed the stellar population in the Orion region
and estimated that the total number of massive stars ($>8M_{\odot}$)
formed there is $\sim$60. The Carina region is located further away 
at a distance of $\sim2.3$ kpc, but hosts a larger population of 
young massive stars. We have estimated the total number to be 
$\sim$ 360 stars ($>8M_{\odot}$). With our population synthesis tool we
calculated the history of kinetic energy, $^{26}$Al and $^{60}$Fe
and UV radiation output from the massive stars in Orion
(\cite{Voss-orion}) and Carina (\cite{Voss-carina}). The results of
applying our population synthesis model to the Orion OB1 association
are shown in figure \ref{fig:1}.

The results were shown to be 
consistent with observed properties of the region: ISM excavation by the 
kinetic energy, free-free emission from the ionized ISM and gamma-rays 
from the radioactive decay of $^{26}$Al. In particular, the results
support massive stellar models with relatively strong wind mass-loss,
and therefore provide evidence against a reduction of mass-loss rates
beyond the revision of 

%and \cite{RefJ}


\begin{thebibliography}{}
\bibitem[Voss et al.(2009)]{Voss-popsyn} 
Voss, R., Diehl, R., Hartmann, D.~H., et al.\ 2009, A\&A, 504, 531 
\bibitem[Voss et al.(2010)]{Voss-orion} Voss, R., Diehl, R., Vink, J.~S., \& Hartmann, D.~H.\ 2010, A\&A, 520, A51
\bibitem[Voss et al.(2011)]{Voss-carina} Voss, R., Martin, P., Diehl, R., Vink, J.~S., Hartmann, D.~H., Preibisch, T., A\&A, submitted
\bibitem[Vink et al.(2000)]{Vink2000} 
Vink, J.~S., de Koter, A., \& Lamers, H.~J.~G.~L.~M.\ 2000, A\&A, 362, 295
\end{thebibliography}
\end{document}